\documentstyle[sprocl,psfig]{article}
\def\Journal#1#2#3#4{{#1} {\bf #2}, #3 (#4)}

\def\NPB{{\em Nucl. Phys.} B}
\def\PLB{{\em Phys. Lett.}  B}
\def\PRL{\em Phys. Rev. Lett.}
\def\PRD{{\em Phys. Rev.} D}
\def\ZPC{{\em Z. Phys.} C}
\def\ZPA{{\em Z. Phys.} A}
\def\NPA{{\em Nucl. Phys.} A}

\def\ra{\rightarrow}

\def\al{\alpha}
\def\be{\begin{equation}}
\def\ee{\end{equation}}
\def\bea{\begin{eqnarray}}
\def\eea{\end{eqnarray}}
\def\rd{{\rm d}}
\def\mev{\,{\rm MeV}}
\def\gev{\,{\rm GeV}}

\newcommand{\das}{distribution amplitudes}

\begin{document}
\thispagestyle{empty}
\hspace*{9.0cm}                        WU-B 97-20\\
\hspace*{10.0 cm}                       July  1997\\            
\\
\begin{center}
{\Large\bf ELECTROMAGNETIC PROCESSES AND THE DIQUARK MODEL \footnote
{Invited talk presented at the Conference on Perspectives in Hadronic Physics,
Trieste (May 1997)}} \\
\vspace*{1.0 cm}
\end{center}
\begin{center}
{\large P. Kroll}\\
\vspace*{0.5 cm}
Fachbereich Physik, Universit\"{a}t Wuppertal, \\
D-42097 Wuppertal, Germany\\[0.3 cm]
\end{center}
\newpage
\setcounter{page}{1}

\title{ELECTROMAGNETIC PROCESSES AND THE DIQUARK MODEL}

\author{PETER KROLL \footnote{e-mail:
kroll@theorie.physik.uni-wuppertal.de\\
Supported in part by the TMR Network ERB 4061 PL 95 0115.}}

\address{Fachbereich Physik, Universit\"at Wuppertal,\\
D-42097 Wuppertal, Germany}
\vspace{0.9cm}
\maketitle\abstracts{
The present status of the diquark model for exclusive 
reactions at moderately large momentum transfer is reviewed. That 
model is a variant of the Brodsky-Lepage approach in which diquarks are
considered as quasi-elementary constituents of baryons. Recent
applications of the diquark model, relevant to high energy physics with
electromagnetic probes, are discussed: electromagnetic form factors of
baryons in both the space-like and the time-like region,
two-photon annihilations into proton-antiproton pairs as well as 
real and virtual Compton scattering.}

\begin{figure}[t]
\vspace*{-0.2cm}
\[
\psfig{figure=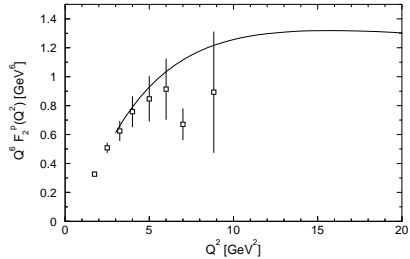,%
        bbllx=0pt,bblly=520pt,bburx=485pt,bbury=830pt,%
        height=3.5cm,clip=}
\]
\vspace*{-0.9cm}
\caption[dummy4]{The Pauli form factor of the proton scaled by
  $Q^6$. Data are taken from \cite{bos:92}. The solid line represents the
  result obtained with the diquark model \cite{jak:93}.}
 \label{fig:pauli}
\vspace*{-0.5cm}
\end{figure}

\begin{figure}[t]
\vspace*{-0.5cm}
\[
    \psfig{figure=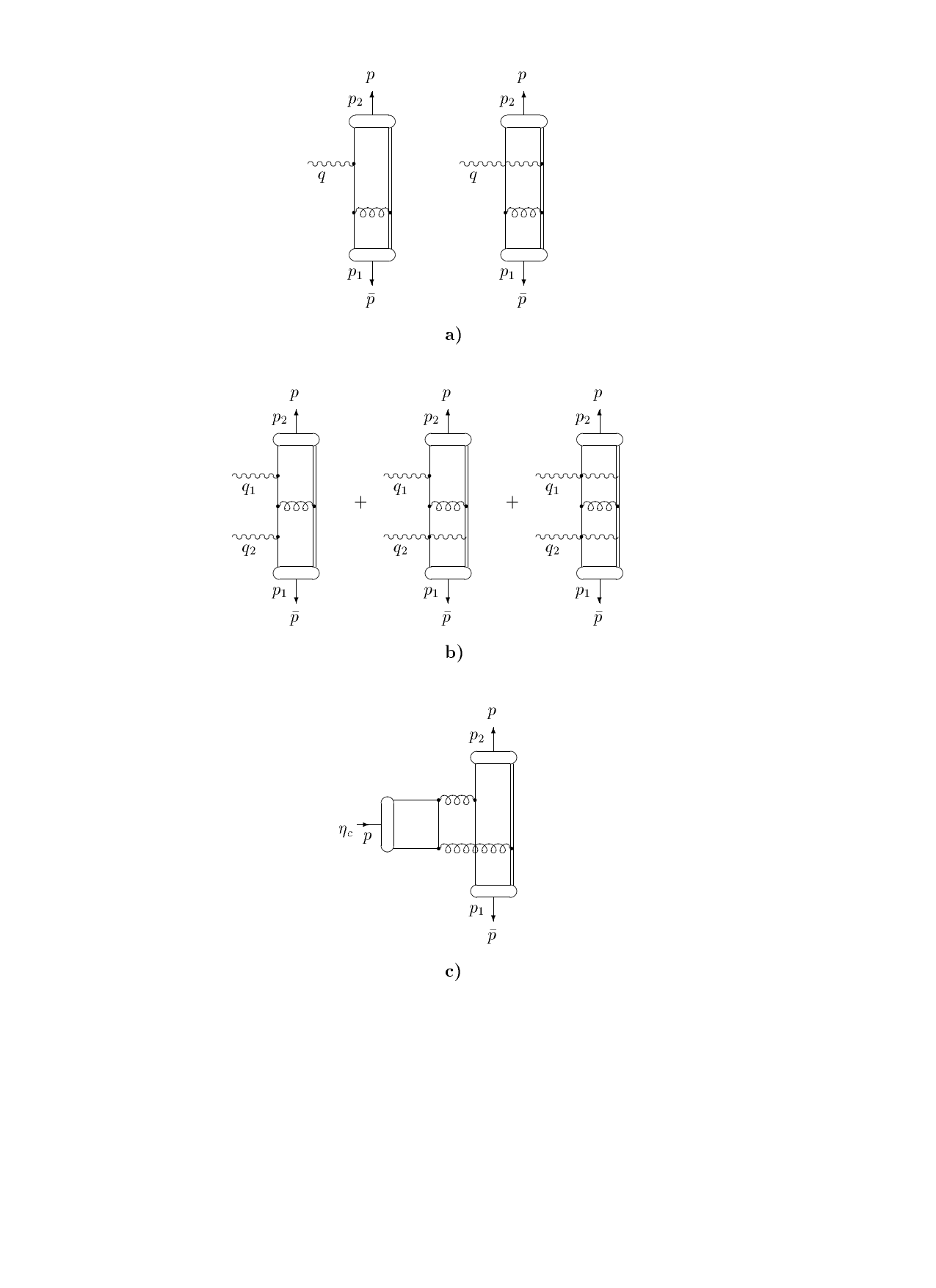,%
        bbllx=140pt,bblly=423pt,bburx=430pt,bbury=600pt,%
        height=3cm,clip=}
\]
\vspace*{-1.0cm}
\caption[]{Typical Feynman graphs contributing to  
$\gamma^{(\ast)}\,p\rightarrow\gamma\,p$.}
\label{frfd}
\end{figure}

\begin{figure}[t]
\[
    \psfig{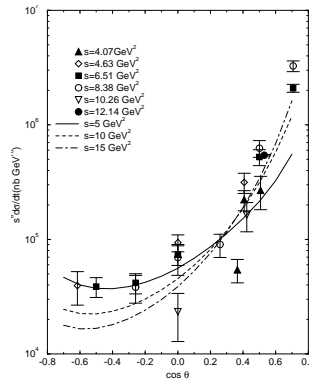}
\]
\vspace*{-1.0cm} 
\caption[]{The scaled cross section for RCS off protons  
vs.\ $\cos\theta$ for three different photon energies. 
The experimental data are taken from \cite{shu:79}.} 
\label{frcc}
\vspace*{-0.5cm} 
\end{figure}

\begin{figure}[t]
\[
    \psfig{figure=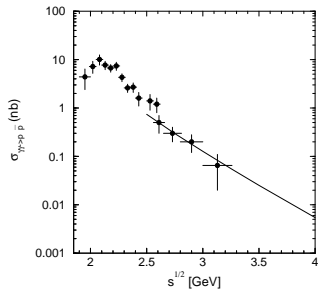,%
        bbllx=70pt,bblly=220pt,bburx=470pt,bbury=573pt,%
        height=4cm,clip=}
\]
\vspace*{-0.9cm}
\caption[]{The integrated $\gamma\gamma\ra p\bar{p}$ cross section
  ($\mid\cos{\theta}\mid\ge 0.6$). The solid line represents the diquark
  model prediction \cite{kro:93a}. Data are taken from CLEO \cite{cleo}.} 
\label{time}
\vspace*{-0.5cm} 
\end{figure}

\begin{figure}[t]
\[
    \psfig{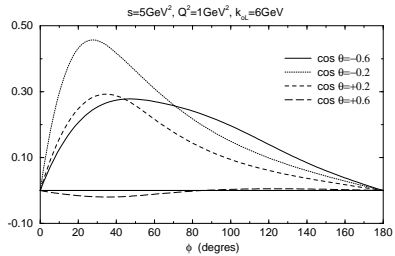}
\]
\vspace*{-1.0cm}
\caption[dummy5] {Diquark model predictions for the electron asymmetry
                  in $ep\to ep\gamma$ \cite{kro:96a}.}
\label{fig:asym}
\vspace*{-0.5cm}
\end{figure}

Exclusive processes at large momentum transfer are described in terms
of hard scatterings among quarks and gluons \cite{lep:80}. In this
so-called hard scattering approach (HSA) a hadronic amplitude
is represented by a convolution of process independent
\das{} (DA) with hard scattering amplitudes to be calculated within
perturbative QCD. The DAs specify the distribution
of the longitudinal momentum fractions the constituents carry. They
represent Fock state wave functions integrated over transverse
momenta. The convolution manifestly factorizes long (DAs) and short
distance physics (hard scattering). The HSA has two characteristic properties,
the power laws and the helicity sum rule. The first property
says that, at large momentum transfer and large
Mandelstam $s$, the fixed angle cross section of a reaction $AB \ra CD$
behaves as (apart from powers of $\log{s}$)
\be
\label{a1}
\vspace*{-0.2cm}
\rd \sigma /\rd t = f (\theta)\, s^{2-n}
\ee
where n is the minimum number of external particles in the hard 
scattering amplitude. The power laws also apply to form factors:
a baryon form factor behaves as $1/Q^4$, a meson form factor as $1/Q^2$.
The counting rules are found to be in surprisingly good agreement 
with experimental data. Even at momentum transfers as low as 2 GeV
the data seem to respect the counting rules.\\
The second characteristic property of the HSA is the conservation of
hadronic helicity. For a two-body process the helicity sum rule reads
\be
\label{a2}
\vspace*{-0.2cm}
\lambda_A + \lambda_B = \lambda_C + \lambda_D.
\ee
It appears as a consequence of utilizing the 
collinear approximation and of dealing with (almost) massless quarks
which conserve their helicities when interacting with gluons.    
The collinear approximation implies that the relative orbital angular 
momentum between the constituents has a zero component in the
direction of the parent hadron. Hence the helicities of the
constituents sum up to the helicity of their parent hadron.
The helicity sum rule is violated by $20-30 \%$ in many cases.
A particular striking example is the Pauli form factor of the proton
which is determined by a helicity flip contribution. 
Its $Q^2$ dependence (see Fig.\ \ref{fig:pauli}) is compatible with 
a higher twist contribution
($\sim 1/Q^6$).\\  
In explicit applications of the HSA (carried through only in valence
quark approximation and to lowest order QCD with very few exceptions) 
one encounters the difficulty that the data are available only at 
moderately large momentum transfer, a region in which non-perturbative
dynamics may still play a crucial role. A general feature of these 
applications is the extreme sensitivity of the results to the DAs 
chosen for the involved hadrons. Only strongly end-point concentrated 
DAs provide results which are in a few cases in fair agreement with 
the data \cite{cer:84}. These apparent successes 
of the HSA are only achieved at the expense of strong contributions 
from soft regions where one of the constituents carries only a tiny 
fraction of its parent hadron's momentum. This is a very problematical
situation for a perturbative calculation. It should be stressed that 
none of the DAs used in actual applications leads to a successful
description of all large momentum transfer processes investigated so far.\\
It seems clear from the above remarks that the HSA (in the valence
quark approximation) although likely to be the correct asymptotic 
picture for exclusive reactions, needs modifications at moderately 
large momentum transfer. In a series of papers \cite{ans:87}$^{-}$\cite{kro:96b} 
such a modification has been proposed by us in which baryons are
viewed as composite of quarks and diquarks. The latter are treated as 
quasi-elementary constituents which partly survive medium hard
collisions. Diquarks are an effective description of correlations in 
the wave functions and constitute a particular model for
non-perturbative effects. The diquark model may be viewed as a variant
of the HSA appropriate for moderately large momentum transfer and it 
is designed in such a way that it evolves into the standard pure quark
HSA asymptotically. In so far the standard HSA and the diquark model 
do not oppose each other, they are not alternatives but rather complements.
The existence of diquarks is a hypothesis. However, from experimental
and theoretical approaches there have been many indications suggesting
the presence of diquarks. For instance, they were introduced in baryon
spectroscopy, in nuclear physics, in astrophysics, in jet fragmentation 
and in weak interactions to explain the famous $\Delta I=1/2$ rule. 
Diquarks also provide a natural explanation of the equal slopes of 
meson and baryon Regge trajectories. For more details and for 
references, see \cite{kro:87}. It is important to note that QCD
provides some attraction between two quarks in a colour $\{\bar{3}\}$ 
state at short distances as is to be seen from the static reduction of the
one-gluon exchange term. \\
Even more important for our aim, diquarks have also been found to play a role in
inclusive hard scattering reactions. The most obvious place to signal
their presence is deep inelastic lepton-nucleon scattering. Indeed
the higher twist contributions, convincingly observed by the NMC 
\cite{vir:91}, can be modelled as lepton-diquark elastic scattering. 
Baryon production in inclusive $pp$ collisions also reveals the need 
for diquarks scattered elastically in the hard interaction
\cite{szc:92}. For instance, kinematical dependences or the excess of
the proton yield over the antiproton yield find simple explanations
in the diquark model. No other explanation of these phenomena is
known as yet.\\
{\bf {\it The diquark model:}} As in the standard HSA a helicity amplitude
for the reaction $AB\ra CD$ is expressed as a convolution of DAs and
hard scattering amplitudes ($s$, $-t$, $-u$\,\, $\gg m_i^2$) 
\begin{eqnarray}
\label{helamp}
{\large M}(s,t)\,=\,\int \rd x_C \rd x_D \rd x_A \rd x_B  
\Phi^*_C(x_C) \Phi^*_D(x_D) T_H(x_i,s,t) \Phi_A(x_A) \Phi_B(x_B) 
\end{eqnarray}
where helicity labels are omitted for convenience. Implicitly
it is assumed in (\ref{helamp}) that the valence Fock states consist
of only two constituents, a quark and a diquark (antiquark) in the
case of baryons (mesons). In so far the specification of the quark
momentum fraction $x_i$ suffices; the diquark (antiquark) carries the
momentum fraction $1-x_i$. If an external particle is point-like,
e.g. a photon, the accompanying DA is to be replaced by $\delta (1-x_i)$.
As in the standard HSA contributions from higher Fock states are
neglected. This is justified by the fact that that such contributions 
are suppressed by powers of $\al_s/t$ as compared to that from the 
valence Fock state.\\
In the diquark model spin $0$ ($S$) and spin $1$ ($V$) colour
antitriplet diquarks are considered. Within flavour SU(3) the $S$ 
diquark forms an antitriplet, the $V$ diquark an sixtet. 
Assuming zero relative orbital angular momentum between quark and
diquark and taking advantage of the collinear approximation, the
valence Fock state of a ground state octet baryon $B$ with helicity
$\lambda$ and momentum $p$ can be written in a covariant fashion 
(omitting colour indices)
\begin{equation}
\label{pwf}
|B;p,\lambda\rangle  = f_S\,\Phi_S^B(x)\,B_S\, u(p,\lambda) 
             + f_V\, \Phi_V^B(x)\, B_V
              (\gamma^{\alpha}+p^{\alpha}/m_B)\gamma_5 \,u(p,\lambda)/\sqrt{3}
\end{equation}
where $u$ is the baryon's spinor. The two terms in (\ref{pwf})
represent configurations consisting of a quark and either a scalar or a 
vector diquark, respectively. The couplings of the diquarks 
with the quarks in a baryon lead to flavour functions which e.g.\ for
the proton read
\vspace*{-0.1cm}
\begin{equation}
\label{fwf}
B_S=u\, S_{[u,d]}\hspace{2cm} 
B_V= [ u V_{\{u,d\}} -\sqrt{2} d\, V_{\{u,u\}}]/\sqrt{3}\, .
\end{equation}
The DAs $\Phi^B_{S(V)}$ are conventionally
normalized as $\int \rd x \Phi = 1$. The constants $f_{S}$ and $f_V$
play the role of the configuration space wave functions at the origin. \\
The DAs containing the complicated non-perturbative bound state physics, 
cannot be calculated from QCD at present. It is still
necessary to parameterize the DAs and to fit the eventual free
parameters to experimental data. Hence, both the models, the standard
HSA as well as the diquark model, only get a predictive power when a
number of reactions involving the same hadrons is investigated. In
the diquark model the following DAs have been proven to work
satisfactorily well in many applications \cite{jak:93}$^{-}$\cite{kro:96b}:
\begin{eqnarray}
\label{a10}
\Phi^B_S(x)&\hspace{-0.3cm}=&\hspace{-0.3cm}N^B_S x (1-x)^3 
                                 \exp{\left[-b^2 (m^2_q/x + m^2_S/(1-x))\right]}\\
\Phi^B_V(x)&\hspace{-0.3cm}=&\hspace{-0.3cm}N^B_V x (1-x)^3 (1+5.8\,x - 12.5\,x^2)
\exp{\left[-b^2 (m^2_q/x + m^2_V/(1-x))\right]}. \nonumber 
\end{eqnarray}
The constants $N^B_S$ and $N^B_V$ are fixed through the normalization  
(e.g.\ for the proton  $N_S^p = 25.97$, $N_V^p = 22.92$). The DAs
exhibit a mild flavour dependence via the exponential whose other purpose 
is to guarantee a strong suppression of the end-point regions. The masses
in (\ref{a10}) are constituent masses; for $u$ and $d$ quarks we take 
$350\mev$ and for the diquarks $580\mev$. Strange quarks and diquarks
are assumed to be $150\mev$ heavier that the non-strange ones. It is 
to be stressed that the quark and diquark masses only appear in the 
DAs (\ref{a10}); in the hard scattering kinematics they are neglected. The final
results (form factors, amplitudes) depend on the actual mass values mildly.
The transverse size parameter $b$ is fixed from the assumption of a
Gaussian transverse momentum dependence of the full wave function and
the requirement of a value of $600\mev$ for the mean transverse
momentum (actually $b = 0.498\gev^{-1}$). As the
constituent masses the transverse size parameter is not considered as
a free parameter since the final results only depend on it weakly.\\ 
The hard scattering amplitudes $T_H$, determined by short-distance
physics, are calculated from a set of Feyman graphs relevant to a
given process. Diquark-gluon and diquark-photon vertices appear in
these graphs which, following standard prescriptions, are defined as 
\begin{eqnarray}
\label{vert}
\mbox{S$\,$g$\,$S}:&& \phantom{-}i\,g_s t^{a}\,(p_1+p_2)_{\mu} \nonumber\\
\mbox{VgV}:&& -i\,g_{s}t^{a}\, 
\Big\{
 g_{\alpha\beta}(p_1+p_2)_{\mu}
- g_{\beta\mu}\left[(1+\kappa)\,p_2-\kappa\, p_1\right]_{\alpha} \nonumber\\
&&\quad\quad\quad - g_{\mu\alpha} \left[(1+\kappa)\,p_1-\kappa\, p_2\right]_{\beta} 
\Big\} 
\end{eqnarray}
where $g_s=\sqrt{4\pi\alpha_s}$ is the QCD coupling constant.
$\kappa$ is the anomalous magnetic moment of the vector diquark and 
$t^a=\lambda^a/2$ the Gell-Mann colour matrix. For the coupling of 
photons to diquarks one has to replace $g_s t^a$ by $-\sqrt{4\pi\alpha} e_D$ 
where $\alpha$ is the fine structure constant and $e_D$ is the electrical 
charge of the diquark in units of the elementary charge. The couplings 
$DgD$ are supplemented by appropriate contact terms required by 
gauge invariance.\\
The composite nature of the diquarks is taken into 
account by phenomenological vertex functions. Advice for the parameterization 
of the 3-point functions (diquark form factors) is 
obtained from the requirement that asymptotically the diquark 
model evolves into the standard HSA. Interpolating smoothly  between 
the required asymptotic behaviour and the conventional value of 1 at $Q^{2}=0$, 
the diquark form factors are actually parametrized as
\begin{eqnarray}
\label{fs3}
\vspace*{-0.5cm}
F_{S}^{(3)}(Q^{2})=\frac{Q_{S}^{2}}{Q_{S}^{2}+Q^{2}}\,,\qquad
F_{V}^{(3)}(Q^{2})=\left(\frac{Q_{V}^{2}}{Q_{V}^{2}+Q^{2}}\right)^{2}\,.
\end{eqnarray}
The asymptotic behaviour of the diquark form factors and the connection to 
the hard scattering model is discussed in more detail in 
Ref.\ \cite{kro:87,kro:91}. In accordance with the required asymptotic
behaviour the $n$-point functions for $n\geq 4$ are parametrized as
\begin{eqnarray}
\label{fsn}
\vspace*{-0.8cm}
F_{S}^{(n)}(Q^{2})=a_{S}F_{S}^{(3)}(Q^{2})\,,\qquad
F_{V}^{(n)}(Q^{2})=
\left(a_{V}\frac{Q_{V}^{2}}{Q_{V}^{2}+Q^{2}}\right)^{n-3}F_{V}^{(3)}(Q^{2}).
\end{eqnarray}
The constants $a_{S,V}$ are strength parameters. Indeed, since the diquarks in 
intermediate states are rather far off-shell one has to consider 
the possibility of diquark excitation and break-up. Both these possibilities 
would likely lead to inelastic reactions. Therefore, we have not to consider 
these possibilities explicitly in our approach but excitation and break-up 
lead to a certain amount of absorption which is taken into account by the 
strength parameters. Admittedly, that recipe is a rather crude
approximation for $n\geq 4$. Since in most cases the contributions from the
n-point functions for $n\geq 4$ only provide small corrections to
the final results that recipe is sufficiently accurate.\\
{\bf {\it Special features of the diquark model:}} The diquark hypothesis
has striking consequences. It reduces the effective number of
constituents inside baryons and, hence, alters the power laws
(\ref{a1}). In elastic baryon-baryon scattering, for instance, the usual power
$s^{-10}$ becomes $s^{-6} F(s)$ where $F$ represents the net effect of
diquark form factors. Asymptotically $F$ provides the missing four
powers of $s$. In the kinematical region in which the diquark model 
can be applied ($-t$, $-u \geq 4 \gev^2$), the diquark form factors 
are already active, i.e.\ they supply a substantial $s$ dependence 
and, hence, the effective power of $s$ lies somewhere between 6 and 10.
The hadronic helicity is not conserved in the diquark model at finite
momentum transfer since vector diquarks can flip their helicities when
interacting with gluons. Thus, in contrast to the standard HSA,
spin-flip dependent quantities like the Pauli form factor of the 
nucleon can be calculated.\\
{\bf {\it Electromagnetic nucleon form factors:}} This is the simplest
application of the diquark model and the most obvious place to fix the
various parameters of the model. The Dirac and Pauli form factors of
the nucleon are evaluated from the convolution formula (\ref{helamp}) 
with the DAs (\ref{a10}) and the parameters are determined from a best
fit to the data in the space-like region. The following set of parameters
\begin{equation}
\label{c1}
\begin{array}{cccc}
 f_S= 73.85\,\mbox{MeV},& Q_S^2=3.22 \,\mbox{GeV}^2, & a_S=0.15, &  \\
 f_V=127.7\,\mbox{MeV},& Q^2_V=1.50\,\mbox{GeV}^2, & a_V=0.05,&\kappa=1.39\,;
\end{array}
\end{equation}
provides a good fit of the data \cite{jak:93}. $\alpha_s$ is evaluated with 
$\Lambda_{QCD}=200\mev$ and restricted to be smaller than $0.5$. The 
parameters $Q_S$ and $Q_V$, controlling the size of the diquarks,
are in agreement with the higher-twist effects observed in the structure 
functions of deep inelastic lepton-hadron scattering \cite{vir:91} if these 
effects are modelled as lepton-diquark elastic scattering. The Dirac
form factor of the proton is perfectly reproduced. The results for the
Pauli form factor are shown in Fig.\ \ref{fig:pauli}. The predictions 
for the two neutron form factors are also in agreement with the data. 
However, more accurate neutron data are needed in the $Q^2$ region of 
interest in order to determine the model parameters better. 
The nucleon's axial form factor \cite{jak:93} and its electromagnetic
form factors in the time-like regions \cite{kro:93a} have also been 
evaluated. Both the results compare well with data. Even 
electroexcitation of nucleon resonances has been investigated 
\cite{kro:92,bol:94}. In the case of the $N\Delta$ form factor the
model results agree very well with recent data \cite{stu96}.\\
{\bf {\it Real Compton scattering (RCS):}} $\gamma p\ra\gamma p$ is 
the next reaction to which the diquark model is applied to. Since the 
only hadrons involved are again protons RCS can be predicted
in the diquark model without any adjustable parameter. Typical
Feynman graphs contributing to that process are shown in Fig.\ \ref{frfd}. 
The results of the diquark model for RCS are shown in Fig.\ \ref{frcc}{}
for three different photon energies \cite{kro:91,kro:96a}. 
Note that in the very forward and backward regions the transverse
momentum of the outgoing photon is small and, hence, the diquark
model which is based on perturbative QCD, is not applicable. Despite
the rather small energies at which data \cite{shu:79} are available, 
the diquark model is seen to work rather well. The predicted
cross section does not strictly scale with $s^{-6}$. The results 
obtained within the standard HSA are of similar quality
\cite{niz:91}. The diquark model also predicts interesting photon 
asymmetries and spin correlation parameters (see the discussion in 
\cite{kro:91}). Even a polarization of the proton, of the order of 
$10\%$, is obtained \cite{kro:91}. This comes about as a consequence 
of helicity flips generated by vector diquarks and of perturbative
phases produced by propagator poles appearing within the domains of
the momentum fraction integrations. The poles are handled in the 
usual way by the $\imath\varepsilon$ presription. The appearance of 
imaginary parts to leading order of $\al_s$ is a non-trivial 
prediction of perturbative QCD \cite{far:89}; it is characteristic 
of the HSA and is not a consequence of the diquark hypothesis.\\
{\bf {\it Two-photon annihilation into $p\bar{p}$ pairs:}} This
process is related to RCS by crossing, i.e.\ the same set of Feynman
graphs contributes (see Fig.\ \ref{frfd}). The only difference is that
now the diquark form factors are needed in the time-like region. 
The expressions (\ref{fs3},\ref{fsn}) represent an effective 
parameterization of them valid at space-like $Q^2$. Since the exact 
dynamics of the diquark system is not known it is not possible to 
continue these parameterizations to the time-like region in a unique way.  
A continuation can be defined as follows \cite{kro:93a}: $Q^2$ is 
replaced by $-s$ in (\ref{fs3},\ref{fsn}) guaranteeing the correct 
asymptotic behaviour and, in order to avoid the appearance
of unphysical poles at low $Q^2$, the diquark form factors are kept
constant once their absolute values have reached $c_0=1.3$ \cite{kro:93a}. 
The same definition of the time-like diquark form
factors is used in the analysis the proton form factor in the
time-like region. The diquark model predictions for the
integrated $\gamma\gamma\ra p\bar{p}$ cross section is compared to the
CLEO data \cite{cleo} in Fig.\ \ref{time}. At large energies the agreement
between predictions and experiment is good. The predictions for the 
angular distributions are in agreement with the CLEO data too. The
standard HSA on the other hand predicts a cross section which lies
about an order of magnitude below the data \cite{far:85}. Recently
CLEO has also measured two-photon annihilations into
$\Lambda\bar{\Lambda}$ pairs \cite{blis:96}. At large energies the
integrated $\Lambda\bar{\Lambda}$ cross section is about a factor of 2
smaller than that one for annihilations into $p\bar{p}$ pairs. 
This may be taken as a hint at a more complicated $\Lambda$ wave
function than (\ref{pwf},\ref{a10}).\\
\newpage
{\bf {\it Virtual Compton scattering (VCS):}} This process is accessible
through $ep \ra ep \gamma$. An interesting element in that reaction 
is that, besides VCS, there is also a contribution from the 
Bethe-Heitler (BH) process where the final state photon is emitted 
from the electron. Electroproduction of photons offers many 
possibilities to test details of the dynamics: One may measure the 
$s$, $t$ and $Q^2$ dependence as well as that on the angle $\phi$ 
between the hadronic and leptonic scattering planes. This allows to
isolate cross sections for longitudinal and transverse virtual
photons. One may also use polarized beams and targets and last but not
least one may measure the interference between the BH and the VC 
contributions. The interference is sensitive to phase differences.\\ 
At $s$, $-t$ and $-u \gg m_p^2$ (or small $|\cos{\theta}|$ where
$\theta$ is the scattering angle of the outgoing photon in the
photon-proton center of mass frame) the diquark model can also be 
applied to VCS \cite{kro:96a}. Again there is no free parameter in 
that calculation. The relevant Feynman graphs are the same as for RCS
(see Fig.\ \ref{frfd}). The model can safely be applied for 
$s\ge 10\gev^2$ and $|\cos{\theta}|\le 0.6$. For the future
CEBAF beam energy of $6\gev$ the model is at its limits of applicability. 
However, since the diquark model predictions for real Compton scattering 
do rather well agree with the data even at $s\ge 5\gev^2$ (see Fig.\ \ref{frcc})
one may expect similarly good agreement for VCS. Predictions for the VCS 
cross section are given in \cite{kro:96a}. The transverse 
cross section (which, at $Q^2=0$, is the cross section for RCS) is
the dominant piece. The other cross sections only become sizeable for
larger values of $|\cos{\theta}|$. Examination of the Bethe-Heitler
contribution to the process $ep\to ep\gamma$ reveals that it is small
as compared to the VCS contribution at high energies, small values of 
$|\cos{\theta}|$ and for an out-of-plane experiment, i.e.\ $\phi\ge 50^{\circ}$.\\
Of interest is also the electron asymmetry in 
$e p \to e p \gamma$:
\begin{equation}
  \label{asy}
A_L\,=\,\frac{\sigma(+) -\sigma(-)}{\sigma(+)+\sigma(+)}  
\end{equation}
where $\pm$ indicates the helicity of the incoming electron. $A_L$
measures the imaginary part of the longitudinal -- transverse 
interference. The longitudinal amplitudes for VCS turn out to be 
small in the diquark model (hence $A_L^{VC}$ is small). However, 
according to the model, $A_L$ is large in the region of strong BH 
contamination (see Fig.\ \ref{fig:asym}). In that region, $A_L$ 
measures the relative phase
(being of perturbative origin from on-shell going internal gluons,
quarks and diquarks \cite{far:89}) between the BH amplitudes
and the VCS ones. The magnitude of the effect shown in
Fig.\ \ref{fig:asym} is sensitive to details of the model and,
therefore, should not be taken literally. Despite of this our results 
may be taken as an example of what may happen. The measurement of 
$A_L$, e.g.\ at CEBAF, will elucidate the underlying 
dynamics of VCS strikingly.\\ 
{\it Summary and outlook:} 
The diquark model which represents a variant of the HSA,
combines perturbative QCD with non-perturbative elements. The diquarks
represent quark-quark correlations in baryon wave functions which are
modelled as quasi-elementary constituents. This model has been applied
to many photon induced exclusive processes at moderarely large
momentum transfer (typically $\geq 4 \gev^2$). From the analysis of
the nucleon form factors the parameters specifying the diquark and the
DAs, are fixed. Compton scattering and two-photon annihilations
of $p\bar{p}$ can then be predicted. The comparison with existing
data reveals that the diquark model works quite well and in fact much
better then the pure quark HSA. \\
Predictions for the VCS cross section and for the $ep\to ep\gamma$ 
cross section have also been made for kinematical situations
accessible at the upgraded CEBAF and perhaps at future high energy 
accelerators like ELFE@HERA. According to the diquark model the BH 
contamination of the photon electroproduction becomes sizeable for 
small azimuthal angles. The BH contribution also offers
the interesting possibility of measuring the relative phases
between the VC and the BH amplitudes. The phases of the VC amplitudes are
a non-trivial phenomenon generated by the fact that some of the internal
quarks, diquarks and gluons may go on mass shell. The electron
asymmetry $A_L$ is particularly sensitive to relative phases.
In contrast to the standard HSA the diquark model allows to
calculate helicity flip amplitudes, the helicity sum rule (\ref{a2})
does not hold at finite $Q^2$. One example of an observable controlled
by helicity flip contributions is the Pauli form factor of the
proton. Also in this case the diquark model accounts for the data.\\
Photo- and electroproduction of mesons can also be calculated within
the diquark model. However, these reactions are already quite 
complicated since all together 158 Feynman graphs contribute. The 
analysis of this class of processes is not yet finished. First results
exist only for $\gamma p\to K^{(*)}\Lambda$ \cite{kro:96b}.\\
The only discrepancy between data and diquark model predictions known
to us, is to be seen in the process
$\gamma\gamma\to\Lambda\bar{\Lambda}$. This failure seems to indicate
that the $\Lambda$ wave function, and perhaps those of other hyperons
are more complicated than (\ref{pwf},\ref{a10}). A solution of this
problem may necessitate a combined analyis of
$\gamma\gamma\to\Lambda\bar{\Lambda}$ and $\gamma p\to K^{(*)}\Lambda$.


\end{document}